\documentclass[aip,cha,preprint,amsmath,amssymb]{revtex4-2}

\usepackage[T1]{fontenc}
\usepackage[utf8]{inputenc}
\usepackage{bm}
\usepackage{graphicx}
\usepackage{booktabs}
\usepackage{siunitx}
\usepackage{xcolor}
\usepackage{hyperref}
\usepackage[nameinlink,noabbrev]{cleveref}

\providecommand{\doi}[1]{\href{https://doi.org/#1}{doi:#1}}

\graphicspath{{./}{figures/}}

\hypersetup{
  colorlinks=true,
  linkcolor=blue!55!black,
  citecolor=blue!55!black,
  urlcolor=blue!55!black
}

\newcommand{\Hcap}{\widehat H_{\rm CAP}}
\newcommand{\ii}{\mathrm{i}}
\newcommand{\Eref}{E_{\rm ref}}
\newcommand{\Eres}{\mathcal E}

\begin{document}

\title{A Phase Space Signature of Quantum Roaming in Chesnavich's Model}

\author{Stephen Wiggins}
\email{stephen.wiggins@me.com}
\affiliation{Hetao Institute of Mathematics and Interdisciplinary Sciences, Shenzhen-Hong Kong International Science Park, No. 3 Binlang Road, Futian Bonded Zone, Shenzhen 518000, China}
\affiliation{School of Mathematics, University of Bristol, Bristol BS8 1UG, United Kingdom}

\date{\today}

\begin{abstract}
Roaming reactions occur when a molecule enters a near-dissociation region, avoids immediate separation, and later forms products by a pathway not controlled by the conventional tight transition-state bottleneck.  Classical studies have shown that roaming is best understood in phase space: inner and outer transition-state structures, together with their invariant manifolds, organize trapping, return, and dissociation.  The corresponding quantum question is less settled.  Can a single quantum resonance carry a recognizable signature of the classical roaming region?  We address this question in Chesnavich's two-degree-of-freedom model for the ion--molecule reaction $\mathrm{CH}_4^+\rightarrow\mathrm{CH}_3^+ + \mathrm{H}$.  Resonance states are computed with a complex absorbing potential and analyzed using diagnostics designed to mirror the classical phase-space picture: radial probability weights derived from the tight and outer transition-state structures, radial Husimi projections, angular-momentum channel weights, and coherent-state probes of the classical periodic orbits.  One resonance is distinguished from the rest of the computed resonance ensemble.  Its wavefunction is concentrated in the projected region between the inner and outer transition-state structures, its radial phase-space distribution is centered at intermediate radius with nearly zero radial momentum, and its angular structure is consistent with a standing rather than a directed rotating component.  We interpret this state as a phase-space-localized quantum analogue of classical roaming.  The result provides a controlled example in which quantum roaming is identified directly from a resonance wavefunction and its phase-space diagnostics, rather than only from product-state or scattering signatures.
\end{abstract}


\maketitle

\begin{quotation}
\noindent\textbf{Chesnavich's model is a minimal Hamiltonian model for an ion--molecule reaction in which a light atom moves relative to a rotating molecular ion.  Despite having only two degrees of freedom, the model contains the essential dynamical ingredients of roaming: an inner tight transition-state structure, an outer loose transition-state structure, and an intermediate region in which trajectories can be trapped before either returning or dissociating.  This article asks whether a quantum resonance can be localized in the same classically defined intermediate region.  The answer is affirmative in the following operational sense: one computed resonance is concentrated between the inner and outer transition-state structures and has radial, angular, and coherent-state phase-space diagnostics consistent with the classical roaming geometry.}
\end{quotation}

\section{Introduction}
\label{sec:introduction}

Transition-state theory organizes chemical reaction dynamics around bottlenecks separating reactants from products.  In the most familiar configuration-space language, such a bottleneck is associated with a saddle point of the potential energy surface.  For a Hamiltonian system, however, the transport object relevant at fixed total energy is not the saddle point by itself.  The saddle is the configuration-space precursor.  Above the saddle energy in a two-degree-of-freedom Hamiltonian system, the corresponding phase-space object is an unstable periodic orbit, born from the equilibrium through the Lyapunov subcenter theorem, together with its associated periodic-orbit dividing surface\cite{wiggins2025phase,MeyerHallOffin2009}.  It is this fixed-energy phase-space structure that controls crossing, recrossing, and flux.

Roaming reactions challenge the idea that reaction proceeds by direct passage through a single compact bottleneck.  In a roaming process, the system begins to dissociate toward radical fragments, enters a broad weakly bound region, avoids prompt separation, and then returns to form molecular products through intramolecular abstraction or related rearrangement.  The trajectory has not simply followed the minimum-energy path through the conventional tight transition-state region.  Instead, it has explored a near-dissociation region in which long-range forces, angular reorientation, and nonstatistical phase-space trapping play a determining role.

The modern study of roaming began with the combined experimental and theoretical identification of the roaming H-atom pathway in formaldehyde decomposition by Townsend and co-workers\cite{Townsend2004}.  Follow-up work established the energy dependence of the formaldehyde roaming pathway\cite{Lahankar2006,Lahankar2007}, and roaming was soon recognized in other unimolecular systems, including acetaldehyde photodissociation\cite{Heazlewood2008}.  Suits emphasized roaming atoms and radicals as a new mechanism in molecular dissociation\cite{Suits2008}, statistical treatments clarified kinetic implications of roaming pathways\cite{Klippenstein2011}, and subsequent reviews placed the phenomenon in a broader chemical context\cite{BowmanShepler2011,BowmanSuits2011,BowmanHouston2017,Suits2020,SuitsOsborn2024}.  These studies show that roaming is no longer a curiosity attached only to formaldehyde.  It is a recurring mechanism in which a near-dissociation complex avoids prompt separation and samples an extended van der Waals or plateau region before forming products.

Acetaldehyde illustrates both the generality and the mechanistic diversity of roaming.  Early work identified roaming as a dominant source of molecular products in acetaldehyde photodissociation\cite{Heazlewood2008}.  More recently, Kraj\v{n}\'ak and Wiggins found evidence for two distinct roaming pathways in acetaldehyde, one at shorter CH$_3$--HCO separation and one at larger separation, with phase-space analysis showing that the shorter-range pathway is not captured by the corresponding reduced-dimensional model\cite{KrajnakWiggins2024}.  This example reinforces a point central to the present paper: roaming should not be identified solely with a large-amplitude excursion in configuration space.  It is controlled by the phase-space structures that organize trapping, return, and dissociation.  Similar phase-space structures have also been used to explain hydrogen-atom roaming in formaldehyde decomposition\cite{Mauguiere2015JPCL}.

Chesnavich's model provides a particularly useful setting for this question.  It was introduced to describe multiple transition states in the ion--molecule dissociation
\begin{equation}
  \mathrm{CH}_4^+\rightarrow \mathrm{CH}_3^+ + \mathrm{H}.
\end{equation}
The model retains a radial coordinate, describing the separation of the hydrogen atom from the methyl cation, and an angular coordinate, describing relative orientation.  It is therefore small enough for detailed classical and quantum analysis, but still rich enough to contain the competition between an inner tight transition-state structure and an outer loose, or orbiting, transition-state structure\cite{Chesnavich1986}.  Ezra and Wiggins later derived the Hamiltonian explicitly as a rigid body coupled to a structureless particle, clarifying the mechanical origin of the kinetic-energy terms and the invariant structures of the model\cite{EzraWiggins2019}.

Classical studies of Chesnavich's model have shown how roaming trajectories can be classified using periodic-orbit dividing surfaces and the invariant manifolds associated with inner and outer periodic orbits\cite{Mauguiere2014CPL,Mauguiere2014JCP,KrajnakWiggins2018,KrajnakWaalkens2018,Mauguiere2017ARPC}.  In two degrees of freedom, these organizing structures are unstable periodic orbits on a fixed energy surface.  The tight transition state (TTS) is the inner transition-state periodic orbit and its dividing surface.  The outer transition state (OTS) is the outer loose or orbiting periodic orbit and its dividing surface.  The FR1 family is the first family of hindered-rotor periodic orbits that appears in the intermediate radial region.  By ``family'' we mean a continuous branch of periodic orbits obtained by continuation in a system parameter, most commonly the total energy.  The term ``hindered rotor'' refers to angular motion in which the relative orientation
coordinate \(\theta\) is not freely rotating but is constrained by the angular potential
barrier in \(V(r,\theta)\). The FR1 family is the first periodic-orbit family associated
with this hindered angular motion in the intermediate radial region. In the present paper, however, all quantum diagnostics are evaluated at a fixed reference energy, so the relevant object is the member of the FR1 periodic-orbit family at that energy and, more specifically, its radial turning-point range.  In the classical roaming picture, the region between the TTS and OTS structures is the phase-space region in which trajectories can undergo roaming-type trapping and return, while the FR1 structure identifies a narrower nonlinear-resonance region nested inside that broader roaming interval.

The corresponding quantum problem is less developed.  The phrase ``quantum roaming'' has several meanings in the existing literature.  One meaning refers to quantum effects, such as tunneling, zero-point motion, or path-integral delocalization, that modify roaming-like complex-forming dynamics at low temperature.  For example, del Mazo-Sevillano and co-workers used ring-polymer molecular dynamics to study low-temperature reactions of OH with formaldehyde and methanol and identified long-lived complex-forming mechanisms associated with quantum roaming\cite{DelMazo2019}.  Another meaning, closer to the present work, concerns quantum resonances, scattering structures, or wavefunctions that bear the imprint of roaming phase-space dynamics.  Li, Li, and Guo described a quantum manifestation of roaming in the H + MgH reaction, including the emergence of roaming resonances associated with large-amplitude motion near threshold\cite{Li2013}.  The subsequent MgH$_2$ study by Maugui\`ere and co-workers combined classical and quantum dynamics on an \emph{ab initio} surface, used phase-space structures to identify trapping and roaming pathways, and observed quantum bound and resonance patterns related to projected periodic orbits\cite{Mauguiere2016JPCA}.  In formaldehyde, detailed product-state correlations have revealed rotational and orbiting resonances coupled to roaming, radical, and molecular channels\cite{Quinn2020,Foley2021,Foley2022}.  From a methodological point of view, quantum roaming can therefore be approached either through time-dependent wavepacket or scattering dynamics, or through metastable resonance states.  We adopt the resonance viewpoint because classical roaming is a trapping-and-escape phenomenon, and resonances are stationary quantum signatures of temporary trapping before dissociation.

In this paper we use the term \emph{quantum roaming} in a deliberately operational sense.  A quantum-roaming resonance is a metastable resonance state whose wavefunction and phase-space diagnostics are localized in the classically defined roaming region between the inner tight transition-state structure and the outer transition-state structure, rather than being localized near the TTS or in the asymptotic outgoing region.  Here ``metastable'' means that the state is temporarily localized in the interaction region for a finite time and is associated with a complex energy whose imaginary part gives the decay width and lifetime.  This definition does not assert that every trajectory-based element of the classical roaming definition carries over directly to quantum mechanics.  It identifies a quantum analogue of the classical phase-space signature.

The main result is the identification of a primary quantum-roaming resonance in Chesnavich's model.  Among the computed complex absorbing potential (CAP) resonances, one state is separated from the others by a consistent set of diagnostics.  Its probability is concentrated in the projected TTS--OTS interval, its radial Husimi distribution peaks at intermediate radius with nearly zero radial momentum, and its angular distribution is dominated by balanced positive and negative angular-momentum components.  Coherent-state probes placed along the classical TTS, FR1, and OTS periodic orbits show stronger local phase-space compatibility with the outer/roaming structure than with the tight transition-state structure.  The conclusion is not that the state is located on the OTS periodic orbit.  Rather, it is a resonance supported in the intermediate region organized by the inner and outer phase-space bottlenecks.

The paper is organized as follows.  \Cref{sec:model} introduces the classical and quantum Chesnavich model, the fixed-energy transition-state structures, the radial probability diagnostics, and the CAP resonance calculation.  \Cref{sec:methods} explains the wavefunction and phase-space diagnostics used to identify quantum roaming.  \Cref{sec:results} presents the primary resonance and its comparison with the computed resonance ensemble.  \Cref{sec:discussion} discusses robustness and interpretation.  \Cref{sec:conclusion} summarizes the conclusions and outlines next steps.

\section{Classical and quantum Chesnavich model}
\label{sec:model}

We use the two-degree-of-freedom Chesnavich Hamiltonian\cite{Chesnavich1986,EzraWiggins2019} in coordinates $(r,\theta)$ with conjugate momenta $(p_r,p_\theta)$:
\begin{equation}
  H(r,\theta,p_r,p_\theta)
  = \frac{p_r^2}{2\mu}
  + \frac{1}{2}\left(\frac{1}{I_{\rm CH_3}}+\frac{1}{\mu r^2}\right)p_\theta^2
  + V(r,\theta).
  \label{eq:classical_H}
\end{equation}
The potential is written
\begin{equation}
  V(r,\theta)=V_{\rm CH}(r)+\frac{1}{2}V_0(r)\bigl(1-\cos 2\theta\bigr),
  \label{eq:potential}
\end{equation}
with
\begin{equation}
  V_{\rm CH}(r)=\frac{D_e}{c_1-6}\left[2(3-c_2)e^{c_1(1-x)}-(4c_2-c_1c_2+c_1)x^{-6}-(c_1-6)c_2x^{-4}\right],
  \qquad x=\frac{r}{r_e},
  \label{eq:VCH}
\end{equation}
and
\begin{equation}
  V_0(r)=V_e\exp[-\alpha(r-r_e)^2].
  \label{eq:V0}
\end{equation}
The Hamiltonian is a reduced mechanical model of a mobile H atom interacting with a rigid symmetric-top representation of CH$_3^+$.  For reproducibility, we state the parameter values used in the present work.  Following the standard Chesnavich/Maugui\`ere parameter set\cite{Chesnavich1986,Mauguiere2014JCP}, we take
\begin{equation}
  D_e=47\,\mathrm{kcal\,mol^{-1}},\quad r_e=1.1\,\text{\AA},\quad c_1=7.37,\quad c_2=1.61,\quad V_e=55\,\mathrm{kcal\,mol^{-1}},
\end{equation}
\begin{equation}
  m_H=1.007825\,\mathrm{u},\quad m_C=12.0\,\mathrm{u},\quad I_{\rm CH_3}=2.373409\,\mathrm{u}\,\text{\AA}^2,
\end{equation}
and
\begin{equation}
  m_{\rm CH_3}=m_C+3m_H,\qquad \mu=\frac{m_Hm_{\rm CH_3}}{m_H+m_{\rm CH_3}}\simeq 0.944467\,\mathrm{u}.
\end{equation}
The parameter value used here is $\alpha=1\,\text{\AA}^{-2}$, corresponding to the late-transition-state case in the Chesnavich model.  The kinetic coupling through
\begin{equation}
  A(r)=\frac{1}{I_{\rm CH_3}}+\frac{1}{\mu r^2}
\end{equation}
allows the angular and radial motions to interact nontrivially.

The classical reference energy used to compute the fixed-energy periodic-orbit reference structures is
\begin{equation}
  \Eref=0.5.
\end{equation}
At this fixed energy, the classical transition-state objects are periodic orbits on the energy surface and their associated periodic-orbit dividing surfaces.  The TTS is the inner unstable periodic orbit associated with the tight transition-state bottleneck.  The OTS is the outer unstable periodic orbit associated with the loose, orbiting bottleneck.  The FR1 family is an intermediate hindered-rotor family.  Since the quantum wavefunction is represented in the $(r,\theta)$ coordinates, we use radial diagnostics derived from these fixed-energy phase-space structures.  More precisely, we project the relevant periodic orbits from phase space onto the radial coordinate and extract either a single radial boundary or a radial turning-point range.  The quantities reported below are therefore not new definitions of the TTS, FR1, and OTS themselves; they are radial diagnostic boundaries derived from the corresponding fixed-energy periodic orbits.  For the reference computation at $\Eref=0.5$, the validation-derived radial turning-point values are
\begin{align}
  r_{\rm TTS,outer} &\simeq 2.116, \nonumber \\
  r_{\rm FR1,min}  &\simeq 3.179, \nonumber \\
  r_{\rm FR1,max}  &\simeq 3.651, \nonumber \\
  r_{\rm OTS}      &\simeq 13.386.
  \label{eq:classical_markers}
\end{align}
The values in \cref{eq:classical_markers} are rounded to physically meaningful plotting and diagnostic precision.  Full numerical precision can be retained in the computational data files, but it is not needed in the main text.  The FR1 interval is much narrower than the full TTS--OTS interval because it is the radial span of a specific fixed-energy member of the FR1 periodic-orbit family, not the whole roaming region.  Its role in the quantum analysis is therefore not to redefine roaming more narrowly, but to provide a nested diagnostic for localization near the nonlinear resonance structure inside the broader TTS--OTS roaming interval.

These classical structures motivate the following radial probability diagnostics for a resonance wavefunction $\psi(r,\theta)$:
\begin{align}
  P_{\rm inner} &= \int_{r<r_{\rm TTS,outer}} |\psi(r,\theta)|^2\,dr\,d\theta, \\
  P_{\rm FR1} &= \int_{r_{\rm FR1,min}\le r\le r_{\rm FR1,max}} |\psi(r,\theta)|^2\,dr\,d\theta, \\
  P_{\rm roam} &= \int_{r_{\rm TTS,outer}\le r\le r_{\rm OTS}} |\psi(r,\theta)|^2\,dr\,d\theta, \\
  P_{\rm outer} &= \int_{r>r_{\rm OTS}} |\psi(r,\theta)|^2\,dr\,d\theta.
  \label{eq:prob_diagnostics}
\end{align}
Here $P_{\rm roam}$ is a radial diagnostic induced by the classical TTS and OTS phase-space structures.  It is not a characteristic function of the full four-dimensional roaming region in phase space.  The FR1 probability is a nested subdiagnostic, since the FR1 radial window lies inside the TTS--OTS interval used in $P_{\rm roam}$.  Thus the four quantities in \cref{eq:prob_diagnostics} should be read as diagnostic probabilities, not as four mutually disjoint bins.  When a disjoint radial partition is desired, one may define
\begin{equation}
  P_{\rm roam}^{\rm excl}=P_{\rm roam}-P_{\rm FR1},
\end{equation}
which is the probability in the TTS--OTS interval excluding the FR1 window.

The quantum resonances are computed using a complex absorbing potential (CAP), a standard finite-domain method for representing outgoing resonance states and extracting resonance energies and widths\cite{RissMeyer1993,Muga2004}.  The CAP-modified Hamiltonian is
\begin{equation}
  \Hcap=\widehat H-\ii\eta W(r),
  \label{eq:H_CAP}
\end{equation}
where $W(r)$ is the CAP profile and $\eta>0$ is the CAP strength.  The profile is chosen to vanish in the physical region and to grow only in an outer absorbing layer.  In the baseline calculation we use a normalized quadratic form,
\begin{equation}
 W(r)=
 \begin{cases}
  0, & r<r_{\rm CAP}, \\
  \left(\dfrac{r-r_{\rm CAP}}{r_{\max}-r_{\rm CAP}}\right)^2, & r_{\rm CAP}\le r\le r_{\max},
 \end{cases}
 \label{eq:cap_profile}
\end{equation}
with
\begin{equation}
  \eta=2.0\times10^{-2}, \qquad r_{\rm CAP}=22, \qquad r_{\max}\simeq34.89.
  \label{eq:cap_parameters}
\end{equation}
The motivation for this choice is simple.  The absorber begins well outside the OTS radius, $r_{\rm OTS}\simeq13.39$, so it does not intrude into the TTS--OTS roaming interval.  The quadratic profile turns on smoothly at $r_{\rm CAP}$ and reaches order one at the edge of the numerical box, reducing artificial reflection from the absorber interface while allowing outgoing amplitude to be removed before it reaches $r_{\max}$.  The CAP strength $\eta$ is a baseline value chosen to identify stable resonance branches without making the absorber so strong that it strongly distorts the interior wavefunction.

CAP eigenvalues are complex.  We write them in the standard resonance form
\begin{equation}
  \Eres_j=E_j-\frac{\ii}{2}\Gamma_j,
\end{equation}
so that
\begin{equation}
  \Gamma_j=-2\,\mathrm{Im}\,\Eres_j.
\end{equation}
The motivation for this convention is the time dependence
\begin{equation}
  \exp(-\ii\Eres_j t/\hbar)
  =\exp(-\ii E_jt/\hbar)\exp[-\Gamma_j t/(2\hbar)].
\end{equation}
The probability therefore decays as $\exp(-\Gamma_j t/\hbar)$, and $\Gamma_j$ is the resonance width.  Since the absolute width can depend on CAP details, the geometric localization diagnostics are the primary evidence for quantum roaming in this study.

The primary resonance reported below has real part $E\simeq0.502260$, whereas the classical periodic-orbit reference structures were computed at $\Eref=0.5$.  This is a relative difference of about $0.45\%$.  The classical structures vary smoothly over such a small energy interval, and here they are used as a reference geometry for the resonance diagnostics.  A final high-precision study over a range of energies should recompute the classical reference structures at the real part of each resonance energy, but this small offset does not affect the qualitative classification made below.

\subsection{Computed resonance ensemble and state indexing}
\label{subsec:ensemble_indexing}

The CAP diagonalization and subsequent diagnostic filtering produced a retained resonance survey of 32 candidate states in the energy window considered.  The integer labels used below, such as state 10 or state 14, are internal indices of this retained computational ensemble.  They are not additional quantum numbers and do not imply an ordering by energy or by physical importance.  Their purpose is simply to identify particular CAP resonance candidates consistently across the diagnostic tables and figures.

The primary resonance is state 10.  It is selected because it combines a large projected roaming probability, small outer-region probability, an intermediate-radius radial Husimi maximum near zero radial momentum, and a stable branch identity under the CAP, box, and grid perturbations reported below.  State 14 is the secondary candidate: it also has substantial projected roaming probability but has much larger outer-region weight.  States 12, 11, 5, 20, 21, and 25 are comparison states from the same retained ensemble.  They are included to show how the primary state differs from mixed or predominantly outer-localized resonances.

\section{Diagnostics for quantum roaming}
\label{sec:methods}

The object analyzed in this section is a CAP resonance eigenfunction $\psi(r,\theta)$ of the quantum Chesnavich Hamiltonian in the two-coordinate representation.  It is normalized for diagnostic purposes before the probability weights and phase-space projections are computed.  A resonance is not an ordinary bound state.  It is a metastable state with a complex energy: it can remain temporarily localized in the interaction region for a finite time before decaying by dissociation into the fragment channel.  In this language, ``decay'' refers to loss of resonance amplitude or probability from the interaction region, while ``dissociation'' refers to the underlying chemical escape to $\mathrm{CH}_3^+ + \mathrm{H}$.  This makes resonances natural objects for quantum roaming.  If roaming corresponds classically to trapping between inner and outer transition-state structures, then a quantum-roaming resonance should show wavefunction and phase-space support in the corresponding region before decay.

The most direct diagnostic is the configuration-space density $|\psi(r,\theta)|^2$ and its radial marginal
\begin{equation}
  \rho(r)=\int |\psi(r,\theta)|^2\,d\theta.
\end{equation}
However, configuration-space density alone does not distinguish localization in position from localization in phase space.  We therefore use three additional diagnostics.

First, we compute a radial Husimi projection.  A Husimi function is a coherent-state phase-space representation of a quantum state\cite{Husimi1940}.  It can be viewed as a positive, Gaussian-smoothed phase-space diagnostic: it does not have the oscillatory sign changes of a Wigner function, and it is well suited for visualizing simultaneous position--momentum localization.  Coherent-state and Gaussian wavepacket methods have long been used in semiclassical dynamics and phase-space visualization\cite{Heller1975,Heller1981,Littlejohn1986}.

For the present problem, a full four-dimensional Husimi representation in $(r,\theta,p_r,p_\theta)$ would be expensive and difficult to visualize.  We therefore use a projected radial diagnostic.  Let
\begin{equation}
  g_{r_0,p_{r,0}}(r)=C\exp\left[-\frac{(r-r_0)^2}{2\sigma_r^2}+\frac{\ii}{\hbar}p_{r,0}(r-r_0)\right]
  \label{eq:radial_coherent_state}
\end{equation}
be a radial coherent state centered at the radial phase-space point $(r_0,p_{r,0})$, with width $\sigma_r$ and normalization constant $C$.  This Gaussian packet is a local test function: it asks whether the resonance has support near a specified radial position and radial momentum.  Projecting the full wavefunction onto this radial coherent state and then integrating over the angular coordinate gives
\begin{equation}
  Q_r(r_0,p_{r,0})=\int \left|\int g_{r_0,p_{r,0}}^*(r)\psi(r,\theta)\,dr\right|^2 d\theta.
  \label{eq:radial_husimi}
\end{equation}
The maximum of $Q_r$ gives a compact diagnostic of the dominant radial phase-space support of the resonance.  In particular, a maximum near $p_r=0$ in the TTS--OTS radial interval indicates slow radial motion in the projected roaming region, not merely large configuration-space density.

Second, we compute angular-momentum channel weights.  Since $\theta$ is an angular coordinate, its conjugate momentum is represented quantum mechanically by
\begin{equation}
  \widehat p_\theta=-\ii\hbar\frac{\partial}{\partial\theta}.
\end{equation}
The Fourier modes $e^{\ii m\theta}$ are therefore angular-momentum eigenchannels, with angular momentum $m\hbar$.  Expanding
\begin{equation}
  \psi(r,\theta)=\sum_m \psi_m(r)e^{\ii m\theta},
\end{equation}
we define
\begin{equation}
  P_m=\int |\psi_m(r)|^2\,dr,
\end{equation}
with the normalization convention inherited from the normalized resonance.  The word ``channel'' here simply denotes one angular Fourier component of the resonance wavefunction.  A state dominated by $m=+m_0$ has a definite sense of angular motion, while a state dominated by $m=-m_0$ has the opposite sense.  If the $+m_0$ and $-m_0$ components are nearly balanced, their angular momenta cancel in the expectation value, giving $\langle p_\theta\rangle\simeq0$.  The corresponding angular pattern is then a standing structure, similar to a cosine-type angular wave, rather than a directed rotating wavepacket.

Third, we use coherent-state probes along the classical TTS, FR1, and OTS periodic orbits.  The motivation comes from semiclassical phase-space analysis and the theory of periodic-orbit localization.  Quantum wavefunctions can display enhanced intensity near classical organizing structures, including unstable periodic orbits; this is the phenomenon of quantum scarring in chaotic systems\cite{Heller1984}.  Here the periodic-orbit probes are used more modestly.  They are a local phase-space microscope for testing whether the resonance is compatible with neighborhoods of the classical TTS, FR1, and OTS structures.

At a point $z_k=(r_k,\theta_k,p_{r,k},p_{\theta,k})$ on a periodic orbit, we define a product coherent state $\chi_{z_k}$ localized near $z_k$ and compute
\begin{equation}
  w_k=|\langle \chi_{z_k}|\psi\rangle|^2.
  \label{eq:coherent_orbit_probe}
\end{equation}
Sampling $w_k$ along an orbit and recording its maximum and period average gives a relative measure of local phase-space compatibility with that periodic-orbit neighborhood.  These weights should not be interpreted as probabilities that the resonance is ``on'' a periodic orbit.  A periodic orbit is a one-dimensional curve in the classical phase space, while a resonance wavefunction is spread over a finite phase-space region.  The useful information is therefore comparative: using the same coherent-state probe construction on the TTS, FR1, and OTS orbits, which classical structure has the strongest local compatibility with the resonance?

\section{Results}
\label{sec:results}

\subsection{Classical radial regions}

\begin{figure}[t]
  \centering
  \includegraphics[width=0.98\textwidth]{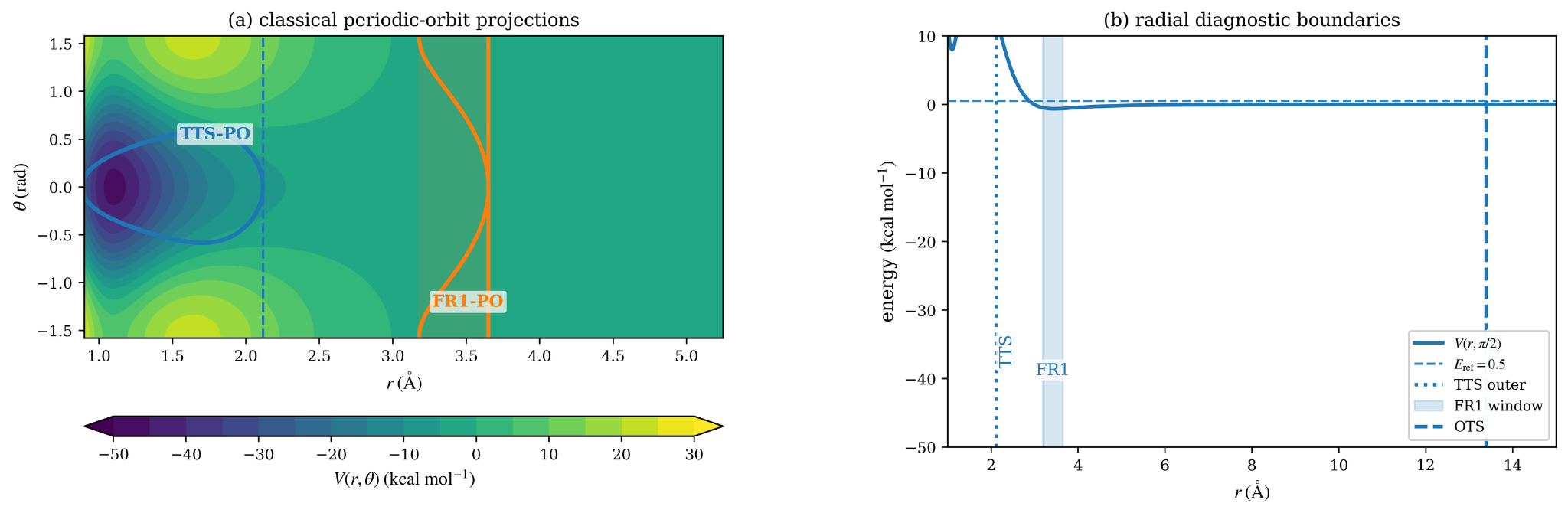}
  \caption{Classical reference structures and radial diagnostic boundaries used to define the projected quantum diagnostics.  Panel (a) shows the Chesnavich potential-energy contours for $\alpha=1$ with representative radial--angular projections of the TTS and FR1 periodic orbits at $\Eref=0.5$.  Panel (b) shows the corresponding radial diagnostic boundaries derived from the fixed-energy periodic-orbit computation.  The wide interval between the TTS and OTS projections is used as a radial diagnostic of the classically defined roaming region, while the FR1 window is a nested subdiagnostic within that interval.}
  \label{fig:classical_regions}
\end{figure}

\Cref{fig:classical_regions} summarizes the classical reference structures used throughout the analysis.  Panel (a) shows the Chesnavich potential-energy contours in the inner radial region together with representative projections of the TTS and FR1 periodic orbits at $\Eref=0.5$.  Panel (b) shows how the same fixed-energy periodic-orbit information is converted into radial diagnostic boundaries: the TTS outer radial boundary, the FR1 turning-point window, and the OTS radius.  The full classical roaming region is a phase-space region organized by the TTS and OTS periodic orbits and their invariant manifolds; the radial intervals in panel (b) are projected diagnostics used to interpret the quantum resonance wavefunctions.

\subsection{Diagnostic summary of the primary resonance}

The primary quantum-roaming resonance is state index 10 in the computed CAP resonance ensemble.  Its complex energy is
\begin{equation}
  E\simeq0.502260,\qquad \Gamma\simeq6.34\times10^{-4}.
\end{equation}
The additional digits used internally for branch tracking are not physically important for the interpretation in this section.  With the corrected validation-derived radial diagnostic boundaries in \cref{eq:classical_markers}, the radial probability diagnostics are
\begin{align}
  P_{\rm inner} &\simeq 0.0073, \\
  P_{\rm FR1} &\simeq 0.0165, \\
  P_{\rm roam} &\simeq 0.9060, \\
  P_{\rm outer} &\simeq 0.0867.
\end{align}
Because $P_{\rm FR1}$ is a nested diagnostic inside $P_{\rm roam}$, the corresponding disjoint roaming probability outside the FR1 window is
\begin{equation}
  P_{\rm roam}^{\rm excl}=P_{\rm roam}-P_{\rm FR1}\simeq0.8895.
\end{equation}
Thus the state has most of its normalized probability in the projected TTS--OTS interval, very little probability in the inner/TTS region, and less than $9\%$ beyond the OTS.  This is the basic probability evidence for classifying the state as projected-roaming-localized rather than tight-transition-state-localized or outer-dissociative.

The radial density peak is at
\begin{equation}
  r_{\rho,\max}\simeq6.86,
\end{equation}
while the radial Husimi maximum is at
\begin{equation}
  r_H\simeq6.89,\qquad p_{r,H}=0.
\end{equation}
The agreement between the radial density maximum and the radial Husimi maximum shows that the phase-space maximum is not an artifact of the coherent-state projection.  The fact that the Husimi maximum occurs at nearly zero radial momentum supports the interpretation of the state as slowly moving, or temporarily trapped, in the projected roaming interval.

\begin{figure}[t]
  \centering
  \includegraphics[width=0.98\textwidth]{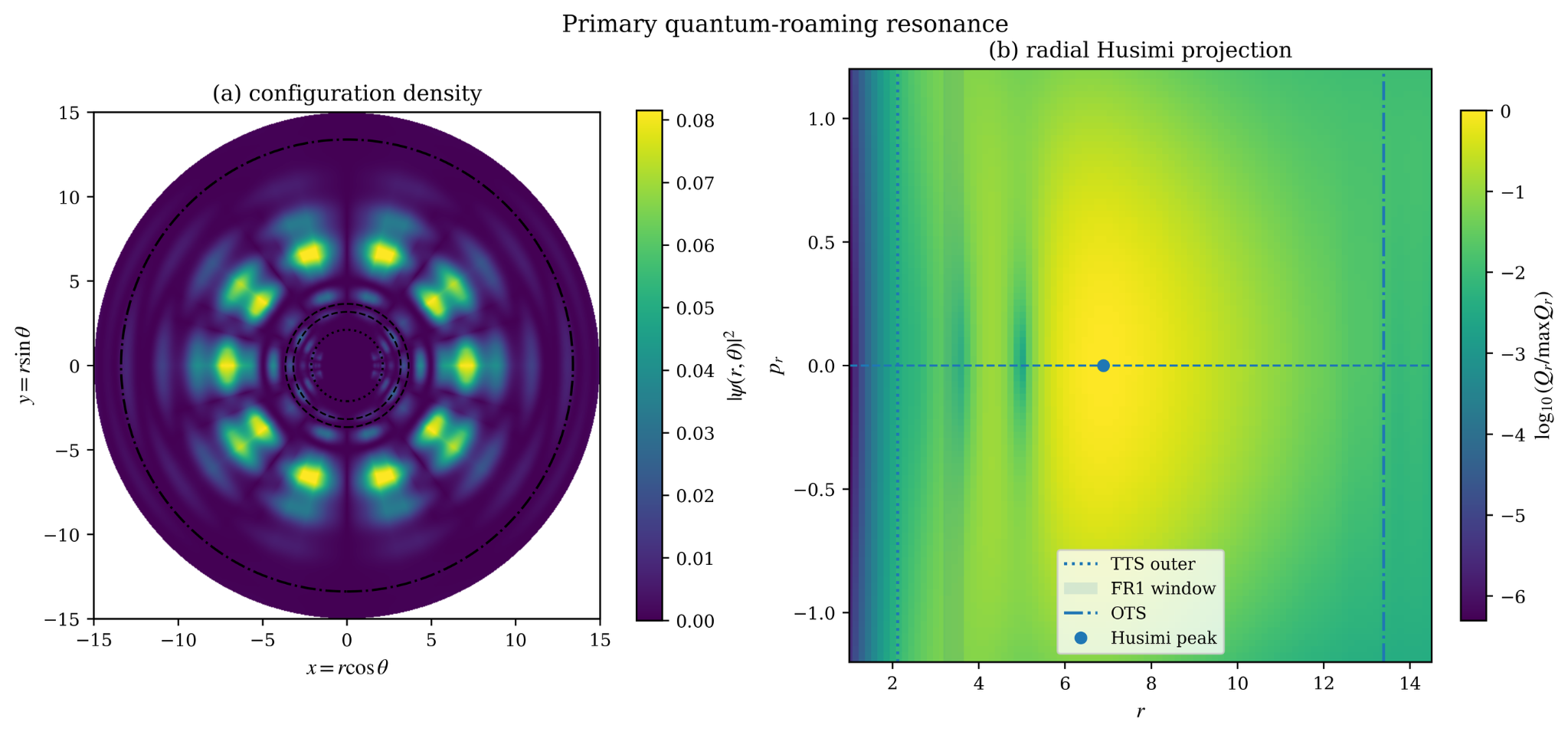}
  \caption{Primary quantum-roaming resonance.  Left: configuration-space density $|\psi(r,\theta)|^2$ with circles marking the TTS outer boundary, the FR1 window, and the OTS radius.  Right: radial Husimi projection in $(r,p_r)$.  The Husimi maximum lies at intermediate radius with nearly zero radial momentum, inside the OTS boundary and outside the tight transition-state region.}
  \label{fig:primary_atlas}
\end{figure}

\Cref{fig:primary_atlas} gives the main diagnostic summary.  The configuration-space density occupies an intermediate radial region and is not concentrated near the inner transition-state structure.  The radial Husimi map shows a compact maximum near $r\simeq6.9$ and $p_r=0$, well inside the OTS projection and far outside the TTS region.

\subsection{Angular and periodic-orbit quantum signatures}

The angular-channel decomposition gives
\begin{equation}
  \langle p_\theta\rangle \simeq 0,
\end{equation}
within numerical precision, and the dominant angular weight is
\begin{equation}
  |m|=5,\qquad P_{|m|=5}\simeq0.858.
\end{equation}
As explained in \cref{sec:methods}, the interpretation is that the $m=+5$ and $m=-5$ components are nearly balanced.  Their opposite angular momenta cancel in the expectation value, so the angular pattern is standing rather than a directed rotating packet.  We do not assign this $|m|=5$ dominance to a specific classical resonance family here.  A systematic study of whether this channel belongs to a larger family of quantum roaming resonances is left for future work.

\begin{figure}[t]
  \centering
  \includegraphics[width=0.96\textwidth]{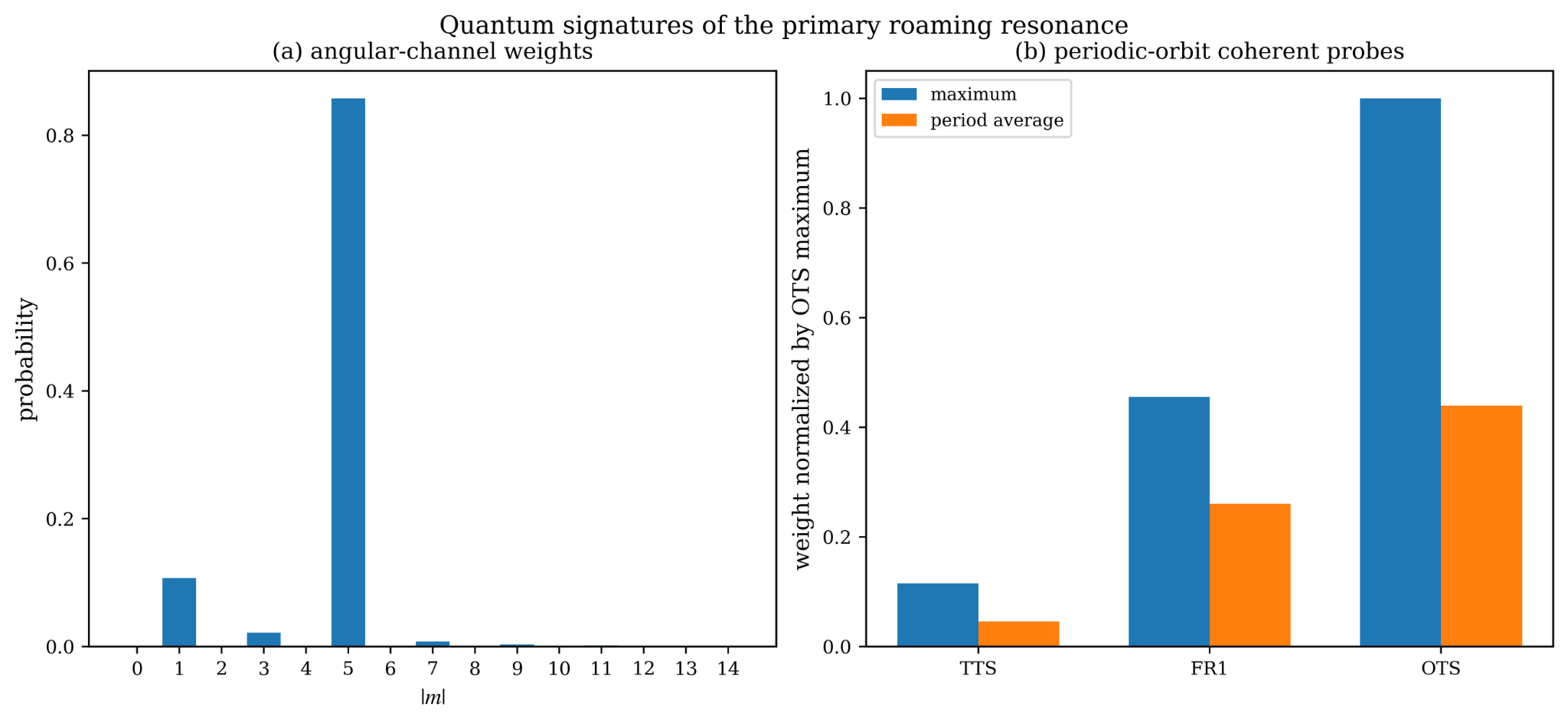}
  \caption{Quantum signatures of the primary resonance.  Left: angular-channel weights, showing dominance of the balanced $|m|=5$ channel.  Right: coherent-state probe weights on the TTS, FR1, and OTS periodic orbits, normalized by the OTS maximum for this state.  These are relative local probe intensities, not probabilities of occupation of the periodic orbits.}
  \label{fig:primary_signatures}
\end{figure}

The coherent-state periodic-orbit probe weights satisfy
\begin{equation}
  W_{\rm OTS}^{\max}\simeq8.33\times10^{-4},\qquad
  W_{\rm FR1}^{\max}\simeq3.79\times10^{-4},\qquad
  W_{\rm TTS}^{\max}\simeq9.60\times10^{-5}.
\end{equation}
The absolute magnitudes are small because each probe is a localized coherent state placed near a point on a one-dimensional periodic orbit, while the resonance is distributed over an extended phase-space region.  These weights are therefore not occupation probabilities.  Their role is comparative.  With the same probe construction applied to the TTS, FR1, and OTS periodic-orbit neighborhoods, the ordering
\begin{equation}
  W_{\rm OTS}>W_{\rm FR1}>W_{\rm TTS}
\end{equation}
indicates stronger local phase-space compatibility with the outer/roaming structure than with the tight transition-state structure.  This supports the roaming interpretation in combination with the probability and Husimi diagnostics; by itself it should not be read as a claim that the resonance is localized on the OTS periodic orbit.  \Cref{fig:primary_signatures} summarizes these angular-channel and coherent-state probe diagnostics.

\subsection{Selection among computed resonances}

\begin{figure}[t]
  \centering
  \includegraphics[width=0.98\textwidth]{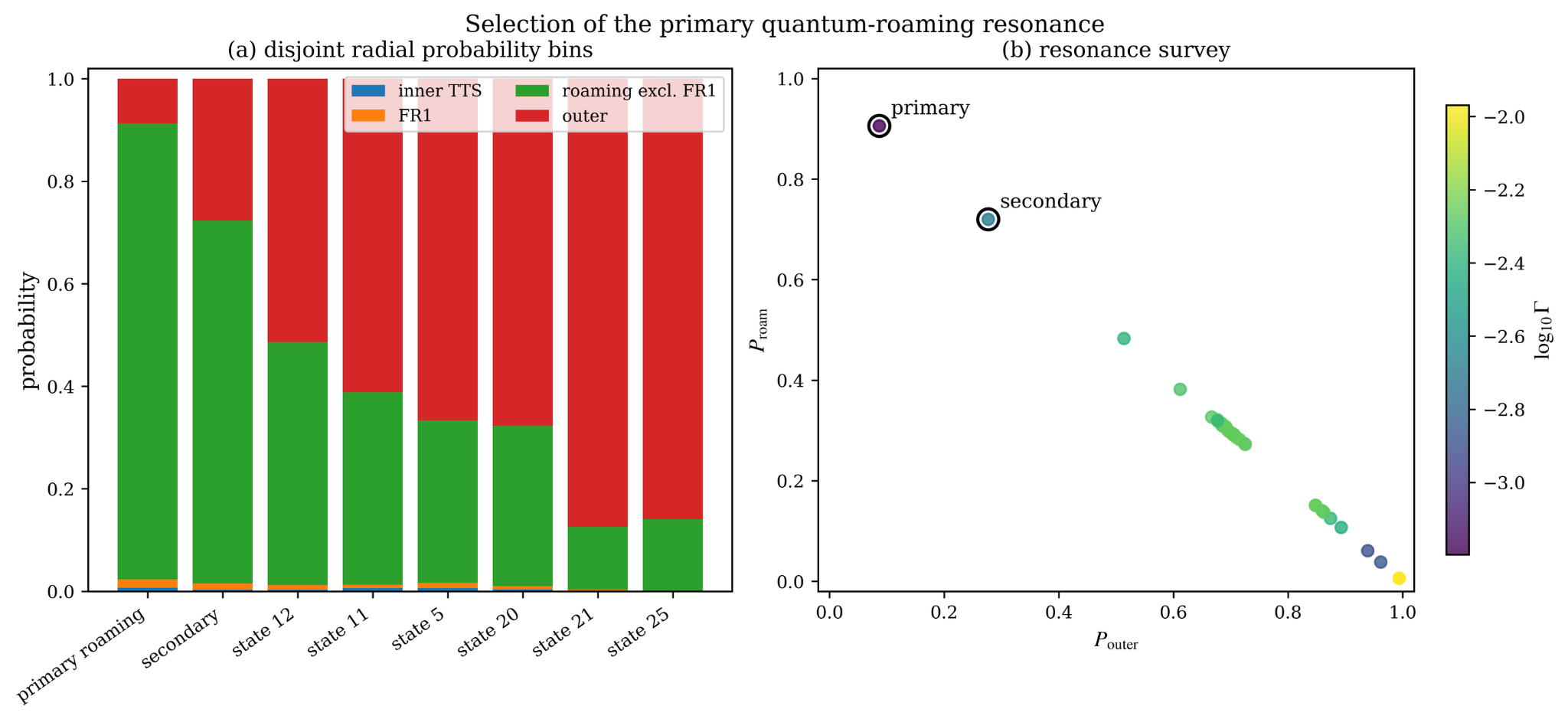}
  \caption{Selection of the primary quantum-roaming resonance from the computed CAP resonance ensemble.  Left: disjoint radial probability display for representative states; the plotted roaming contribution excludes the nested FR1 window so that the displayed bars are additive.  Right: distribution of computed resonances in the $(P_{\rm outer},P_{\rm roam})$ diagnostic plane, with color indicating the logarithm of the CAP width.  The primary state is distinguished by high projected roaming probability and low outer probability.}
  \label{fig:selection}
\end{figure}

The primary state is not chosen by visual inspection alone.  It is isolated from the other computed resonances by the radial probability diagnostics and by the radial Husimi localization.  \Cref{fig:selection} compares representative states and shows the distribution of computed resonances in the $(P_{\rm outer},P_{\rm roam})$ diagnostic plane.  The primary state lies near the top-left of this plane: large projected roaming probability and small outer probability.  The secondary state, state index 14, also has substantial projected roaming probability, but it has much larger outer probability,
\begin{equation}
  P_{\rm roam}\simeq0.720,\qquad P_{\rm outer}\simeq0.277.
\end{equation}
It is therefore better interpreted as a mixed roaming/outer resonance than as the primary quantum-roaming state.

A scalar sorting score was used only as a compact way of ranking candidates.  The top two scores were
\begin{equation}
  S_{10}\simeq0.858,
  \qquad
  S_{14}\simeq0.484,
\end{equation}
while the next state had score near zero.  The gap in this ranking is consistent with the visual separation in \cref{fig:selection}.

\subsection{Radial-density comparison}

\begin{figure}[t]
  \centering
  \includegraphics[width=0.86\textwidth]{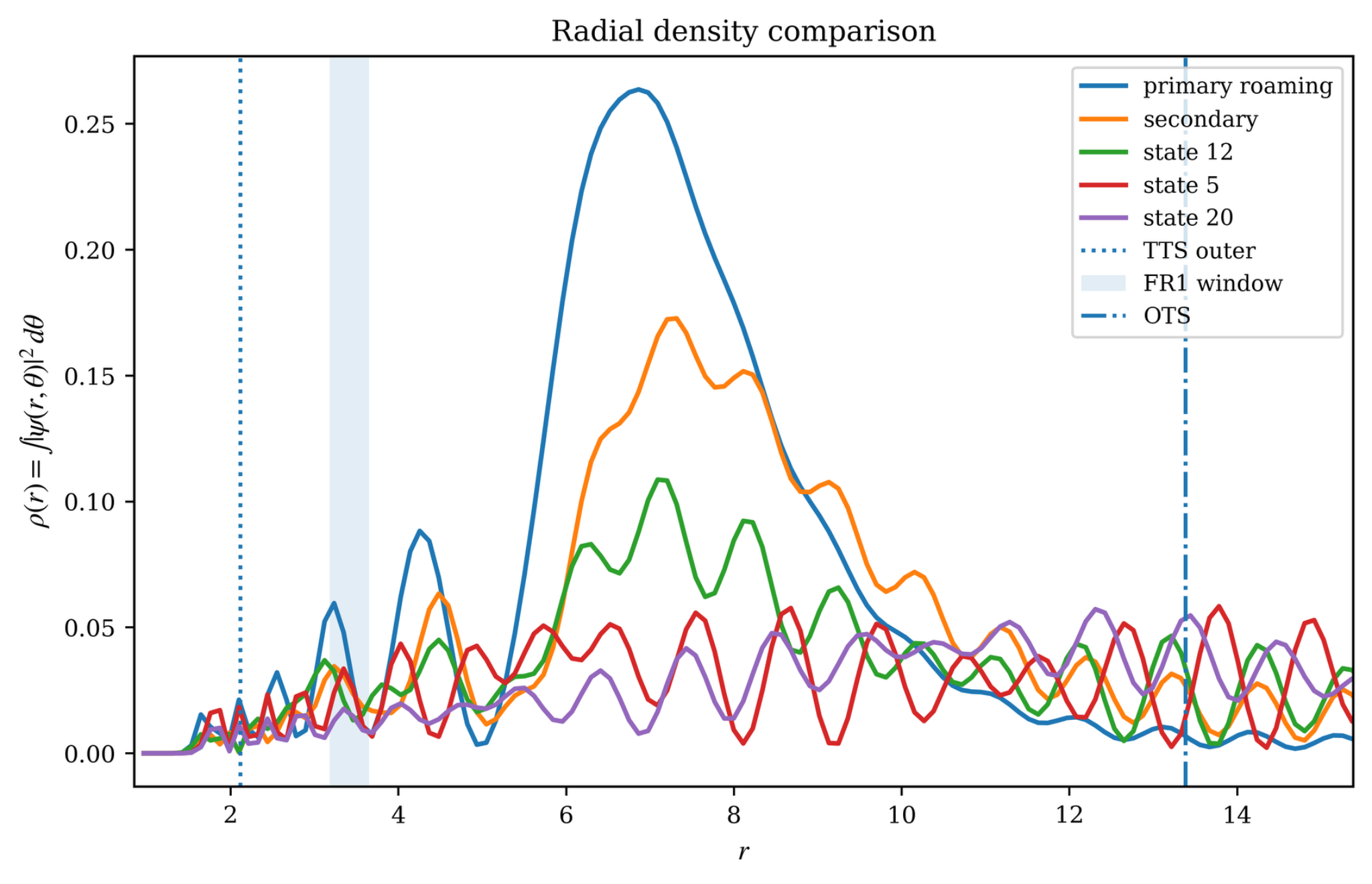}
  \caption{Radial density comparison for the primary resonance, the secondary resonance, and representative control states.  The primary state peaks near $r\simeq6.86$ and remains primarily within the projected TTS--OTS roaming interval defined by the validation-derived radial boundaries.  Control states have substantially larger outer-region weight or less clean localization.}
  \label{fig:radial_density}
\end{figure}

\Cref{fig:radial_density} compares radial marginal densities for the primary, secondary, and representative control states.  The primary state has a strong intermediate-radius peak and relatively low probability beyond the OTS.  Several control states have much stronger support in the outer region and larger CAP widths.  This comparison reinforces the main conclusion: the primary resonance is not merely a broad outer state but a state concentrated in the projected region associated with classical roaming.

\section{Robustness and interpretation}
\label{sec:discussion}

To assess numerical stability, the resonance computation was repeated for 13 runs: one baseline calculation and perturbations of the CAP strength, CAP onset, outer box size, and radial grid resolution.  The baseline used
\begin{equation}
  \eta=2.0\times10^{-2},\qquad r_{\rm CAP}=22,
  \qquad r_{\max}=34.8865,
  \qquad N_r=300,\qquad N_\theta=48,
\end{equation}
with the normalized quadratic CAP profile in \cref{eq:cap_profile}.  The CAP strength was varied over
\begin{equation}
  \eta=1.0\times10^{-2},\ 1.5\times10^{-2},\
  2.5\times10^{-2},\ 3.0\times10^{-2};
\end{equation}
the CAP onset was varied over
\begin{equation}
  r_{\rm CAP}=21.0,\ 21.5,\ 22.5,\ 23.0;
\end{equation}
the box endpoint was varied over
\begin{equation}
  r_{\max}=33.8865,\ 35.8865;
\end{equation}
and the radial grid was varied over
\begin{equation}
  N_r=290,\ 310,
\end{equation}
with $N_\theta=48$ fixed.  For the primary branch, over these 13 runs, the mean and standard deviation of the main geometric diagnostics were
\begin{align}
  P_{\rm roam} &= 0.8941\pm0.0191,\\
  P_{\rm outer} &= 0.0977\pm0.0192,\\
  r_H &= 6.8997\pm0.1689.
\end{align}
The radial Husimi momentum at the grid maximum remained at the grid point nearest zero, $p_{r,H}=-1.74\times10^{-2}$ for this scan.  The branch-overlap diagnostic gave
\begin{equation}
  |\langle\phi_{\rm ref}|\phi_{\rm run}\rangle|^2=0.9740\pm0.0433,
\end{equation}
with minimum value $0.8530$.  These values indicate that the branch identity and the spatial/phase-space localization are stable under the tested perturbations.  The CAP width varied over the scan, with mean $7.18\times10^{-4}$ and standard deviation $1.98\times10^{-4}$ for the primary branch, as expected for an absorber-dependent resonance width.  The robust quantities are therefore the resonance branch and its geometric localization, reported together with the baseline CAP convention.

The results support the following interpretation.  The primary resonance is a metastable quantum state supported in the region between the tight and outer transition-state structures.  It is not an eigenstate of a local well in the ordinary configuration-space sense, nor is it a state localized on a single periodic orbit.  Rather, it is a resonance whose probability is concentrated in the projected radial interval where classical trajectories can roam, whose radial momentum is near zero, and whose angular structure is dominated by a balanced standing component.  The coherent-state periodic-orbit probes further indicate that the state is more compatible with the outer/roaming periodic-orbit neighborhood than with the tight transition-state neighborhood.

This interpretation is consistent with the phase-space view of classical roaming.  In the classical theory, roaming is not simply passage through a special saddle point.  It is controlled by the geometry of the energy surface and by invariant manifolds associated with multiple bottlenecks.  The quantum state identified here appears to inherit that organization: its support is not concentrated at the tight transition state, and it is not absorbed as a purely outgoing outer state.  Instead, it occupies the intermediate region bounded by the relevant phase-space structures.  The interpretation also connects with earlier quantum and mixed classical--quantum roaming studies of H + MgH/MgH$_2$, where resonance wavefunctions displayed patterns related to classical periodic orbits\cite{Li2013,Mauguiere2016JPCA}.  The present contribution differs in that the resonance is selected and interpreted directly by diagnostic probabilities and Husimi/coherent-state localization tied to the classically defined TTS--OTS roaming interval of Chesnavich's model.

It is useful to distinguish this claim from stronger claims that are not made here.  We do not claim that the state is localized on the OTS periodic orbit.  The radial density and Husimi maximum occur near $r\simeq6.9$, substantially inside $r_{\rm OTS}\simeq13.39$.  We also do not claim that the CAP width is an intrinsic observable independent of numerical convention.  The robust object is the resonance branch and its phase-space localization, not the absolute CAP width.  Finally, we do not claim that this one resonance exhausts the meaning of quantum roaming.  The present calculation provides a controlled example in which a metastable quantum state can be tied directly to the classical phase-space roaming region.

\section{Conclusions}
\label{sec:conclusion}

We have identified a primary quantum-roaming resonance in the two-degree-of-freedom Chesnavich model.  The state is distinguished by a mutually consistent set of diagnostics: high probability in the projected TTS--OTS interval, low probability beyond the OTS, a radial Husimi maximum at intermediate radius with nearly zero radial momentum, a balanced angular-momentum structure dominated by $|m|=5$, and stronger coherent-state compatibility with outer/roaming periodic-orbit probes than with the tight transition-state probe.  The state remains geometrically stable under the CAP, box, and grid perturbations tested.

These results support the view that quantum roaming can be identified not only through product-state signatures or scattering observables, but also through direct phase-space diagnostics of individual resonance states.  The key idea is to define quantum diagnostics using classical phase-space structures and then test whether a resonance localizes in the corresponding projected roaming region.  This provides a bridge between classical roaming theory and quantum resonance analysis.

Several extensions are natural.  First, the same analysis should be applied over a range of energies and coupling parameters to determine whether the primary resonance belongs to a family or branch of roaming resonances.  Second, the classical periodic-orbit reference structures should be recomputed at the real part of each resonance energy in such a parameter study.  Third, CAP resonance states should be compared with alternative resonance methods, such as complex scaling or stabilization, to separate geometric localization from absorber dependence.  Fourth, a direct comparison with scattering observables would clarify how the resonance identified here manifests in measurable product distributions.  Finally, the diagnostic strategy developed here could be transferred to higher-dimensional roaming systems, where configuration-space pictures become less reliable and phase-space localization becomes even more important.

\section*{Author Declarations}

\subsection*{Conflict of Interest}
The author has no conflicts to disclose.

\subsection*{Author Contributions}
Stephen Wiggins: Conceptualization, methodology, software, investigation, formal analysis, visualization, and writing--original draft.

\section*{Data Availability}
The numerical data and scripts used to generate the figures are available from the author upon reasonable request.  The figures included in this draft were generated from the corrected final paper-figure workflow, in which the CAP eigenvectors were retained and the radial post-processing diagnostics were recomputed using the validation-derived TTS, FR1, and OTS radial boundaries reported in \cref{eq:classical_markers}.

\begin{acknowledgments}
The author thanks collaborators and colleagues in chemical reaction dynamics and nonlinear dynamical systems for discussions on roaming, phase-space transport, and quantum resonances. 
\end{acknowledgments}

\end{document}